\documentclass[11pt]{article}
\usepackage{amsfonts, amsmath, amssymb, amsthm}
\usepackage[T1,T2A]{fontenc}
\usepackage[utf8]{inputenc}
\usepackage[russian,english]{babel}

\PassOptionsToPackage{obeyspaces}{url}
\usepackage[colorlinks=true,citecolor=blue,urlcolor=blue,linkcolor=blue,bookmarksopen=true]{hyperref}
\usepackage{breakurl}
\usepackage{tikz}

\usepackage{etoolbox} 
\patchcmd{\thebibliography}{\leftmargin\labelwidth}{\leftmargin\labelwidth\addtolength\itemsep{-0.1\baselineskip}}{}{}

\usepackage{mathtools}

\author{Noga Alon
\thanks{Sackler School of Mathematics and Blavatnik School of
Computer Science, Tel Aviv University, Tel Aviv 69978, Israel
and CMSA, Harvard University, Cambridge, MA 02138, USA.  Email:
{\ttfamily nogaa@tau.ac.il}.  
Research supported in part by a BSF grant, an ISF
grant and a GIF grant.}
\and Boris Bukh
\thanks{Department of Mathematical Sciences, Carnegie Mellon University,
Pittsburgh, PA 15213, USA. Email: {\ttfamily bbukh@math.cmu.edu}.
Supported in part by Sloan Research Fellowship
and by U.S.\ taxpayers through NSF CAREER grant DMS-1555149 and NSF grant
DMS-1301548. Part of the work was done during the visit to Universit\'e
Paris-Est Marne-la-Vall\'ee supported by LabEx B\'ezout (ANR-10-LABX-58).}
\and Yury Polyanskiy
\thanks{Department of Electrical Engineering 
and Computer Science, MIT, Cambridge, MA 02139 USA.
	\mbox{Email:~{\ttfamily yp@mit.edu}.} Supported in part by the NSF under Grants No CCF-13-18620 and CCF-17-17842 and by the Center for Science of Information (CSoI),
an NSF Science and Technology Center, under grant agreement CCF-09-39370.}}

\title{List-decodable zero-rate codes}
\date{}

\newtheorem{theorem}{Theorem}
\newtheorem{lemma}[theorem]{Lemma}
\newtheorem{corollary}[theorem]{Corollary}
\newtheorem{proposition}[theorem]{Proposition}

\newtheorem*{definition}{Definition}
\theoremstyle{remark}

\newcommand*{\eqdef}{\stackrel{\text{\tiny{def}}}{=}}            
\newcommand*{\abs}[1]{\lvert #1\rvert}                           
\newcommand*{\norm}[1]{\lVert #1\rVert}                          
\newcommand*{\veps}{\varepsilon}                                 
\newcommand*{\R}{\mathbb{R}}                                     
\newcommand*{\Z}{\mathbb{Z}}                                     
\newcommand*{\E}{\mathop{{}\mathbb{E}}}                          

\def\Ldecodable/{$\mathord<L$-list-decodable}                    
\def\threedecodable/{$\mathord<3$-list-decodable}                %
\DeclareMathOperator{\maxcode}{maxcode}                          
\DeclareMathOperator{\type}{type}                                
\DeclareMathOperator{\rad}{rad}                                  
\DeclareMathOperator{\mrad}{mrad}                                
\DeclareMathOperator{\diam}{diam}                                
\DeclareMathOperator{\ones}{\mathbf{1}}                          
\DeclareMathOperator{\lnk}{lnk}                                  
\DeclareMathOperator{\sign}{sign}                                  
\newcommand*{\Unif}{\mathcal{U}[L]}                              
\newcommand*{\Esp}{\mathcal{H}}                              	
\def\mla{\langle}
\def\mra{\rangle}

\newcommand*{\doi}[1]{\href{http://dx.doi.org/#1}{doi:#1}}

\begin{document}
\maketitle
\begin{abstract}
We consider list-decoding in the zero-rate regime for two cases: 
the binary alphabet 
and the spherical codes in Euclidean space. Specifically, we study the maximal $\tau \in [0,1]$ for which there exists 
an arrangement of $M$ balls of relative Hamming radius $\tau$ in 
the binary hypercube 
(of arbitrary dimension) with the property that no point of the
latter is covered by $L$ or more of them. As $M\to \infty$ the maximal $\tau$ decreases to a well-known critical value
$\tau_L$. In this work, we prove several results on the rate of this convergence. 

For the binary case, we show that the rate is $\Theta(M^{-1})$ when $L$ is even, thus extending the classical
results of Plotkin and Levenshtein for $L=2$. For $L=3$ the rate is shown to be $\Theta(M^{-\tfrac{2}{3}})$. 

For the similar question about spherical codes, we prove the rate is $\Omega(M^{-1})$ and $O(M^{-\tfrac{2L}{L^2-L+2}})$.

\end{abstract}
\section{Introduction}\label{sec:intro}

This work  concerns list-decoding under \emph{worst-case} errors in
the zero-rate regime. We consider the case of the binary
alphabet in Sections~\ref{sec:intro}-\ref{sec:proof_three}  and the case of the unit sphere in Hilbert space in
Section~\ref{sec:hilbert}. 

To motivate our study, we note that the maximum possible size of a code with given parameters is the 
most fundamental combinatorial problem in coding theory. Although the
positive rate region is more interesting for applications, the range
just above the threshold between positive rate and zero rate is intriguing and leads to challenging 
combinatorial problems, exhibiting, as we show in the paper, some interesting behavior.

Specifically, suppose we transmit a sequence of $n$ symbols
from $\{0,1\}$ over a channel that can adversarially change less than
fraction~$\tau$ of the symbols. 
The locations of corrupted symbols are unknown
to the receiver. The goal is to find a code $C\subset \{0,1\}^n$ such
that the receiver can always produce a list of fewer than $L$ messages
containing the transmitted message.  In other words, we seek a code~$C$
such that for every $w\in \{0,1\}^n$ there are fewer than~$L$ codewords
within Hamming distance~$\tau n$ from $w$. 
We call such code \emph{\Ldecodable/} with radius $\tau$. The largest such $\tau$ is denoted by $\tilde \rho_L(C)$ and
is called the \emph{(normalized) $L$-radius} of the code.

Let
\begin{equation}\label{eq:tauL}
  \tau_L\eqdef \frac{1}{2}-\frac{\binom{2k}{k}}{2^{2k+1}}\quad\text{ if }L=2k\text{ or }L=2k+1.
\end{equation}
%
It is known \cite{blinovsky_orig} that for radius $\tau<\tau_L$ the largest \Ldecodable/ code
is exponentially large in $n$, and for radius $\tau>\tau_L$ the largest \Ldecodable/ code is of bounded size (independent of $n$). 
The aim of this paper is to understand how this constant varies 
as $\tau$ approaches $\tau_L$ from above. We define
\[
  \maxcode_L(\veps)\eqdef \max \bigl\{ \abs{C} : C\subset\{0,1\}^n \text{ is \Ldecodable/ of radius }\tau_L+\veps\bigr\}.
\]
Note that in this definition we do not restrict the 
block length~$n$. The maximum is over all choices of $n\in \mathbb{N}$.

We are aware of three results on $\maxcode_L(\veps)$. First, a construction due to Levenshtein \cite{levenshtein} shows that the so-called Plotkin bound is sharp
in the unique decoding case, namely
\[
  \maxcode_2(\veps)=\frac{1}{4\veps}+O(1).
\]
Levenshtein's construction uses Hadamard matrices, and so requires $\veps$ to be of a special form. As a part of Theorem~\ref{thm:simpleconstruction} 
below we present a construction
without a condition on~$\veps$.

Second, Blinovsky \cite{blinovsky_orig} proved that $\maxcode_L(\veps)$
is finite for every $L$ and every $\veps>0$. His proof iterates Ramsey's
theorem, and gives a very large bound on $\maxcode_L(\veps)$ (which is not
made explicit in the paper). Finally, in \cite[Theorem~1]{blinovsky_wrong}
Blinovsky claims an upper bound on $\maxcode_L(\veps)$ of the form
$\maxcode_L(\veps)=O(1/\veps)$. Below we construct a counterexample to
this claim for~$L=3$.\footnote{The mistake appears to stem from the second paragraph of the proof of~\cite[Theorem
1]{blinovsky_wrong}, which proposes a certain extension procedure for codes and claims that it does not decrease $L$-radius. A simple counter-example to the claim is a code $C=\{0,1\}^2$ with $4$-radius equal to 1,
but its extension results in reduction of the $4$-radius to $\tfrac{1}{2}$.} 

We next overview our results for the binary alphabet. The results for spherical codes are in Section~\ref{sec:hilbert}.

\subsection{Our results (binary alphabet)}
Our first result is a version of Levenshtein's construction for any fixed $L$. In comparison to Levenshtein's result, we have no
restriction on $\veps$, but our codes are longer (the value of $n$ is larger).
\begin{theorem}\label{thm:simpleconstruction}
Let $L\geq 2$, suppose $m$ is a positive integer, and let $M=\binom{2m}{m}$. Let $c_L=2^{-L}\lfloor L/2\rfloor\binom{L-1}{\lfloor L/2\rfloor}$. Then there exists
an \Ldecodable/ code in $\{0,1\}^M$ of size~$2m$ and radius $\tau=\tau_L+c_L/2m+O(m^{-2})$. In particular,
\[
  \maxcode_L(\veps)\geq c_L\veps^{-1}+O(1).
\]
\end{theorem}
\begin{theorem}\label{thm:even}
Let $L\geq 2$ be even. Then
\[
  \maxcode_L(\veps)=O(\veps^{-1})
\]
\end{theorem}
We believe that in fact $\maxcode_L(\veps)=c_L\veps^{-1}+O(1)$ for even $L$. 

The case of odd $L$ appears to be significantly harder: the principal reason for this is
Lemma~\ref{lem:tau}(\ref{tauunique}) below. By a different method, however, we were able to make progress for $L=3$:
\begin{theorem}\label{thm:lthree}
We have $\maxcode_3(\veps)=\Theta(\veps^{-3/2})$.
\end{theorem}

\section{Overview of the proof of Theorem~\ref{thm:even}}
The bulk of the paper is devoted to the proof of Theorem~\ref{thm:even}. In this section, we explain
the general structure of the argument.

A code with $\tilde \rho_L(C) = \tau$ is, equivalently, a code in which all $L$-tuples of distinct codewords have large
circumscribed radius, where the latter is defined as the radius of the smallest ball (in Hamming space) containing the
$L$-tuple. First of all, we show (Prop.~\ref{prop:rounding}) that the center of the ball can as well be sought after in  $[0,1]^n$, instead of
$\{0,1\}^n$. The resulting quantity is denoted $\rad(x)$, cf.~\eqref{eq:defrad}. The resulting linear relaxation is much easier to analyze.

Second, we introduce mean radius $\mrad_\omega(x)$ of an $L$-tuple $x$ with respect to a measure $\omega$ on
$\{1,2,\dotsc,L\}$. The significance of this quantity is that there exists (Lemma~\ref{lem:finiteness}) 
finitely many $\omega$'s such that
\begin{equation}\label{eq:rad_intro}
		\rad(x) = \max_{\omega \in \Omega'_L} \mrad_\omega(x) \quad \forall x \in (\{0,1\}^n)^L\,. 
\end{equation}	
Among these $\omega$'s a special one is $\Unif$ -- a uniform distribution on $[L]$.

The mean radii of random independent bits is denoted as $\tau_{\omega,p} = \E[\mrad_\omega(X_1,\ldots,X_L)]$, where $X_i
\stackrel{i.i.d.}{\sim} \mathrm{Ber}(p)$. Properties of $\tau_{\omega,p}$ are summarized in Lemma~\ref{lem:tau}, where
two crucial ones are that $\tau_L = \tau_{\Unif,1/2}$ and that for 
even $L$ (!) we have $\tau_{\omega,p} \le \tau_L -
\delta_0$ for some $\delta_0>0$ and all $\omega \neq \Unif, \omega \in \Omega'_L, p \in [0,1]$.

To understand why $\maxcode_\epsilon(L) = \theta({1\over \epsilon})$ is a natural guess, we recall a standard averaging
argument (Corollary~\ref{cor:meanmrad}):
\begin{equation}\label{eq:mmrad_intro}
		{1\over {M \choose L}} \sum_{x \in {C \choose L}} \mrad_\omega(x) \le \tau_{\omega, 1/2} + O(1/M)\,. 
\end{equation}	
Thus, averaged over the code the mean radii (with respect to any $\omega$) can never exceed $\tau_L + O(1/M)$. Our proof
shows that (for even $L$) and every 
sufficiently large code with $\tilde\rho_L(C) \ge \tau_L$ averaging mean radii
(with $\omega = \Unif$) is equivalent to averaging the actual radius. 

With these preparations our proof proceeds as follows:
\begin{enumerate}
\item We start with showing that a biased code (i.e. one in which 
the fraction of $1$'s among all codewords is $\le 1/2 -
\delta_0$) with radius $\tilde \rho_L(C) \ge \tau_L$ can have at most finite size. This is the content of
Lemma~\ref{lem:smallrad}, which is proved by appealing to a Ramsey theorem to extract a large subcode with the property
that $\rad(x) = \mrad_\omega(x)$ for every $L$-tuple $x$ and some fixed $\omega$. Invocation of a biased-version
of~\eqref{eq:mmrad_intro} then shows this subcode cannot be too large.
\item Next, we show (Lemma~\ref{lem:subsetbias2}) that there is a dichotomy: 
either a code has a large biased subcode
(thus contradicting above), or all but $O(1/M)$ fraction of its $L$-tuples 
must have random-like coordinate composition
(i.e. out of $n$ coordinates about $2^{-L}n$ have 
zeros in all $L$ codewords, and the same holds for each binary pattern). 
In particular, this implies that 
every such $L$-tuple has $\mrad_\omega(x) \approx
\tau_{\omega,1/2}$.
\item Finally, since every $\omega \neq \Unif$ yields $\mrad_\omega(x) \approx \tau_{\omega,1/2} < \tau_L$ we must have
that $1-O(1/M)$ fraction of $L$-tuples satisfy $\rad(x) = \mrad_{\Unif}(x)$ and hence in averaging~\eqref{eq:mmrad_intro} we
can replace $\mrad_\omega(x)$ with $\rad(x)$ and conclude that $\rad(x)$ of a typical $L$-tuple is at most $\tau_L +
O(1/M)$ as claimed.
\end{enumerate}

We mention that our proof of Theorem~\ref{thm:even} also shows why Theorem~\ref{thm:lthree} is perhaps surprising: for
$L=3$ we construct a code such that averaged over the codebook $\mrad_\omega(x)$ is always $\le \tau_L + O(1/M)$ (as it
should be, yet the $\rad(x)$ of every $L$-tuple is $\ge \tau_L + O(1/M^{2/3})$. 
The proof of Theorem~\ref{thm:lthree} relies
on a special relation for radii of triangles in $\ell_1$-spaces, see Prop.~\ref{prop:triangles}.

\section{Mean radii}
\paragraph{Definitions} 
For $x\in \R^n$, let $\norm{x}\eqdef (1/n)\sum \abs{x_i}$. 
In particular, for $x,y\in\{0,1\}^n$
the quantity $\norm{x-y}$ is the 
(normalized) Hamming distance between bit strings $x$ and~$y$.


For points $x^{(1)},\dotsc,x^{(L)}\in \{0,1\}^n$ let 
\begin{equation}\label{eq:defrad}
  \rad(x^{(1)},\dotsc,x^{(L)})=\min_{y\in [0,1]^n} \max_i \norm{x^{(i)}-y}.
\end{equation}
Note that we allow the coordinates of $y$ to be arbitrary real numbers between $0$ and~$1$.
For example, for $C=\{000,100\}$ we have $\rad(C)=1/6$, but for every $y\in \{0,1\}^3$
one of the points of $C$ is at distance at least~$1/3$.
However, this relaxation makes only slight effect, as the next proposition shows.
\begin{proposition}
\label{prop:rounding}
Let $x^{(1)},\dotsc,x^{(L)}\in \{0,1\}^n$. 
If $\tau=\rad(x^{(1)},\dotsc,x^{(L)})$, 
then there is a point $y\in \{0,1\}^n$
such that $\norm{x^{(i)}-y}\leq \tau+\frac{L}{2n}$ for all~$i$.
\end{proposition}
\begin{proof}
For any bit $ z \in \{0,1\}$ and real $w \in [0,1]$ define $\ell(z,w)=w$
if $z=0$ and $\ell(z,w)=1-w$ if $z=1$. Note that with this notation,
for every $i$, $\norm{x^{(i)}-y} =\frac{1}{n} \sum_{j=1}^n \ell
(x^{(i)}_j,y_j)$ is an affine function of the variables $y_j$.

The assumption  $\rad(x^{(1)},\dotsc,x^{(L)})\leq \tau$ is equivalent
to the fact that the polytope in the variables $y_j$ defined by
the inequalities $0 \leq y_j \leq 1$ for all $1 \leq j \leq n$ and
$\frac{1}{n} \sum_{j=1}^n \ell (x^{(i)}_j,y_j) \leq \tau$  for all
$1 \leq i \leq L$  is nonempty. Hence it contains a vertex $y'=(y_1',
\ldots ,y_n')$. In this vertex there are at most  $L$ variables  $y_j'$
which are neither $0$ nor  $1$, and the desired result is obtained by
rounding each such $y_j'$ to the closest integer $y_j$ and by taking
$y_j=y_j'$ for all other coordinates $y_j'$.
\end{proof}

Let $\Omega_L$ be the set of all probability measures on the set~$[L]\eqdef \{1,2,\dotsc,L\}$.
Suppose $\omega\in \Omega_L$ is a probability distribution on $[L]$. 
Then for an $L$-tuple $x=(x^{(1)},\dotsc,x^{(L)})\in (\{0,1\}^n)^L$ of codewords, we define their \emph{mean radius} with respect to $\omega$ by
\begin{equation}\label{eq:defmeanrad}
  \mrad_{\omega}(x)\eqdef \min_{y\in [0,1]^n} \E_{i\sim \omega} \norm{x^{(i)}-y}.
\end{equation}

Because $\E_{i\sim \omega} \norm{x^{(i)}-y}$ can be written as a sum over the individual coordinates,
the $y$ attaining minimum in \eqref{eq:defmeanrad} may be taken to have 
all of its coordinates in $\{0,1\}$.
This leads to an alternative formula for $\mrad_{\omega}(x)$:
\begin{align}\label{eq:meanrad}
  \mrad_{\omega}(x)&=\E_{j\in [n]} \min\left( \sum_{x^{(i)}_j=0} \omega(i), \sum_{x^{(i)}_j=1} \omega(i) \right)\\
\label{eq:meandiff}
                 &=\tfrac{1}{2}-\tfrac{1}{2}\E_{j\in [n]} \left\lvert \sum_{x^{(i)}_j=0} \omega(i)- \sum_{x^{(i)}_j=1} \omega(i) \right\rvert.
\end{align}

\paragraph{Duality} From the comparison of \eqref{eq:defrad} and \eqref{eq:defmeanrad} it is clear that $\rad(x)\geq \mrad_{\omega}(x)$ for any~$\omega$. The key observation is that
a suitable converse holds as well.
\begin{lemma}
For every $x=(x^{(1)},\dotsc,x^{(L)})\in (\{0,1\}^n)^L$ we have
\begin{equation}\label{eq:radmrad}
  \rad(x)=\max_{\omega\in \Omega_L} \mrad_{\omega}(x),
\end{equation}
where the maximum is over all probability measures~$\omega$ on $\{1,2,\dotsc,L\}$.
\end{lemma}
\begin{proof}
Notice that~\eqref{eq:defrad} can be rewritten as
$$ \rad(x) = \min_{y\in [0,1]^n} \max_{\omega \in \Omega_L} \E_{i \sim \omega} \norm{x^{(i)}-y}.
$$
Since the function
	$$ (y, \omega) \mapsto \E_{i \sim \omega} \norm{x^{(i)}-y}\,,$$
is convex in $y$ and affine in $\omega$, von Neumann minimax theorem
~\cite{JVN37} implies
$$ \min_{y\in [0,1]^n} \max_{\omega \in \Omega_L} \E_{i \sim \omega} \norm{x^{(i)}-y} = 
	\max_{\omega \in \Omega_L} \min_{y\in [0,1]^n}  \E_{i \sim \omega} \norm{x^{(i)}-y}\,. $$
Comparing with~\eqref{eq:defmeanrad} completes the proof.
\end{proof}

\begin{lemma}\label{lem:finiteness}
For every $L$ there exists a finite set of probability measures $\Omega_L'\subset \Omega_L$ such that
\begin{equation}\label{eq:radfin}
  \rad(x)=\max_{\omega\in \Omega_L'} \mrad_{\omega}(x)\qquad\text{for all }x\in(\{0,1\}^n)^L.
\end{equation}
Furthermore, $\abs{\Omega_L'}\leq 4^{L^2}$.
\end{lemma}
\begin{proof}
Let $x\in (\{0,1\}^n\bigr)^L$ be any $L$-tuple of words. To each coordinate $j\in [n]$ we can then associate the bit string
\[
  T(j)\eqdef \bigl(x^{(1)}_j,x^{(2)}_j,\dotsc,x^{(L)}_j\bigr).
\]
For a bit string $T\in \{0,1\}^L$, put $P_T=\{j\in [n]: T(j)=T\}$. 

Let $y$ be a point that achieves minimum in \eqref{eq:defrad}. 
For each $T$, replace coordinates of $y$ indexed by $P_T$ by their average. This does not change 
$\norm{x^{(i)}-y}$ and so the obtained point also achieves minimum in \eqref{eq:defrad}. So, we may
assume that $y_j$ depends only on $T(j)$. Denote by $y_T$ the common value of $y_j$ for $j\in T$.
  
Let $\alpha_T=\abs{P_T}/n$. We have that
\[
  \norm{x^{(i)}-y}=\sum_{T\in \{0,1\}^L}\alpha_T \abs{y_T-T_i}.
\]

To each probability measure $\omega\in \Omega_L$ we can associate
a \emph{signature}, which is a function $S_{\omega}\colon \{0,1\}^L\to \{1,-1\}$ defined by
\[
  S_{\omega}(T)\eqdef \operatorname{sign} \left(\sum_{i:T_i=0} 
\omega_i - \sum_{i:T_i=1} \omega_i \right)\qquad\text{for }T\in\{0,1\}^L.
\]

Regard $\omega\to S_{\omega}$ as a linear function on the $\Omega_L$ (which we can identify with a simplex in $\R^{L-1}$).
Since $2^L$ hyperplanes partition $\R^{L-1}$ 
into at most $\sum_{j\leq L-1} \binom{2^L}{j}\leq 2^{L^2}$ regions,
the number of possible signatures is at most $2^{L^2}$.
For each possible signature $S$, let 
$\Omega_S\eqdef\{\omega\in \Omega_L : S_{\omega}=S\}$. 
Since $\Omega_S$ is an intersection of
halfspaces, which, in addition to $S_{\omega}=S$, 
includes the additional inequalities $\omega_i \geq 0$ 
for all $i$ and $\sum_i
\omega_i=1$, it is a convex polytope.

By \eqref{eq:meanrad}, 
the maximum of $\mrad_{\omega}(x)$ over all probability 
measures $\omega\in \Omega_S$ is the maximum of the following linear 
function in the variables $\omega_i$:
$$
\sum_T \alpha_T f(\omega,T)
$$
where 
$$
f(\omega,T) =
\sum_{i:T_i=0} \omega_i\qquad\text{if }S(T)=-1
$$
and
$$
f(\omega,T) =
\sum_{i:T_i=1} \omega_i\qquad\text{if }S(T)=+1.
$$
This maximum is attained at a vertex of the polytope $\Omega_S$. 
Thus, in view of the preceding lemma, we may take $\Omega_L'$
to be the union of the vertex sets of all polytopes $\Omega_S$, 
for all signatures~$S$.

Each $\Omega_S$ is defined by $m\eqdef L+2^L$ inequalities, and so by McMullen's upper bound theorem has at most 
$\binom{m-\lceil L/2\rceil}{\lfloor L/2\rfloor}+\binom{m-\lfloor L/2\rfloor-1}{\lceil L/2\rceil-1}\leq 2^{L^2}$ vertices.
Multiplying by the $2^{L^2}$ possible signatures, we obtain the result.
\end{proof}

The preceding proof gives an algorithm to compute the set $\Omega_L'$. The results of this computation for small $L$ can be found at
\url{http://www.borisbukh.org/code/listdecoding17.html}. Interestingly, for $L\leq 4$ the result is very nice.
For a set $R\subset [L]$ let $\mrad_R(x)$ be the $\mrad_{\omega}(x)$ for the probability measure $\omega$ that is uniform on $R$, i.e.,
$\omega_i=1/\abs{R}$ if $i\in R$. Then
\begin{equation}
\label{eq:smallL}
  \rad(x)=\max_{\abs{R} \text{ is even}} \mrad_R(x)\qquad\text{if }L\leq 4\text{ and }x\in(\{0,1\}^n)^L.
\end{equation}

Our proof of Theorem~\ref{thm:lthree} will use~\eqref{eq:smallL} with $L=3$ and so we establish it formally (generalized
to arbitrary $\ell_1$-vectors).
\begin{proposition}\label{prop:triangles}
For any set of three vectors $x,y,z$ in $\mathbb{R}^n$ with respect to the
$\ell_1$-norm, $\rad(x,y,z)=\tfrac{1}{2}\diam(x,y,z)$.
\end{proposition}
\begin{proof}
Put $x=(x_1,x_2, \ldots ,x_n)$, $y=(y_1,y_2,\ldots ,y_n)$, $z=(z_1,z_2,
\ldots ,z_n)$. Let $d$ be the diameter of the set 
$\{x,y,z\}$. For each $i$ let $m_i$ be the median of
$x_i,y_i,z_i$ and define $m=(m_1,m_2, \ldots ,m_n)$.  Put, also,
$a=\|x-m\|,b=\|y-m\|,c=\|z-m\|$, where $\|\cdot \|$ is the
$\ell_1$-norm.  Note that, crucially $\|x-y\|=a+b, \|y-z\|=b+c,
\|x-z\|=a+c$. Thus each of these three sums is at most $d$.
If each of the quantities $a,b,c$ is at most $d/2$ then $m$ is a
center of an $\ell_1$-ball of radius $d/2$ containing $x,y,z$,
showing that in this case the radius is indeed $d/2$, as needed.
Otherwise one of the above, say $a$, is larger than $d/2$. In this
case define
$
w=(1-\frac{d}{2a})x+\frac{d}{2a} m.
$
It is easy to check that $x-w=\frac{d}{2a}(x-m)$ and hence 
$\|x-w\|=\frac{d}{2a} a=d/2$. In addition
$m-w=(1-\frac{d}{2a})(m-x)$ and hence $\|m-w\|=a-\frac{d}{2}$.

Thus, by the triangle inequality,  
$\|y-w\| \leq \|y-m\|+\|m-w\| =b+a-\frac{d}{2} \leq \frac{d}{2},$
and similarly  $\|z-w\| \leq \frac{d}{2}$, completing the proof.
\end{proof}

\section{Averaging}
Averaging arguments play a major 
role in this paper. We collect them in this section.

\paragraph{Mean radii of random bit strings} Averaging arguments will allow us to show
that, in a large code~$C$, mean radii of codewords from $C$ rarely exceed the mean radius of
random bit strings.  Because of that, we start by computing the mean radius of random bit strings
for arbitrary probability measure~$\omega$. In particular, we will see that the radius threshold 
$\tau_L$ defined in \eqref{eq:tauL} is the average radius of a random $L$-tuple of bit strings.

Call a random string $w\in \{0,1\}^n$ \emph{$p$-biased} if each bit of $w$ is $1$ with probability $p$
and $0$ with probability $1-p$, and the bits are independent of each other. 
For a probability measure $\omega$ on $[L]$, we let 
\[
  \tau_{\omega,p}\eqdef\E \mrad_{\omega}(w^{(1)},\dotsc,w^{(L)}) \qquad \text{for independent $p$-biased }w^{(1)},\dotsc,w^{(L)}\in \{0,1\}^n.
\]
For brevity, write $\tau_{\omega}$ in lieu of $\tau_{\omega,1/2}$. Note that $\tau_{\omega,p}$ is independent of $n$, in view of \eqref{eq:meanrad}.

Let $\Unif$ denote the uniform probability measure on $[L]$.
\begin{lemma}\label{lem:tau}\hspace{2ex}
\begin{enumerate}
\item \label{tauextreme}
For every probability measure $\omega$ on $[L]$ and every $p$ we have
\[
  \tau_{\omega,p}\leq \tau_{\Unif,p}.
\] 

\item \label{tauunique} If $L$ is even and $0<p<1$, then the equality holds if and only if $\omega=\Unif$.


\item \label{taucompute} $\tau_{\Unif,1/2}=\tau_L$, where $\tau_L$ is defined in \eqref{eq:tauL}.

\item \label{taubiased} We have $\tau_{\omega,p}<\tau_L$ whenever $p\neq \tfrac{1}{2}$.
\end{enumerate}
\end{lemma}
\begin{proof}
\textbf{Part \eqref{tauextreme}:}
Given an $\omega$ and a vector of signs $\veps=(\veps_1,\dotsc,\veps_L)\in \{1,-1\}^L$,
define signed sum $f_{\omega}(\veps)\eqdef \sum_i \veps_i \omega(i)$. 
By \eqref{eq:meandiff}, maximization of $\tau_{\omega,p}$ is equivalent to minimization of
\[
  \E_{\veps} \abs{f_{\omega}(\veps)}\qquad\text{for $p$-biased }\veps\in\{1,-1\}^L,
\]
i.e. where $\Pr[\veps_i = -1] = p$.

Let $\Omega^*_p$ be the set of all probability measures on $[L]$ that maximize $\tau_{\omega,p}$. 
(The maximum is achieved because $\omega\mapsto \tau_{\omega}$ is a continuous function on a compact set.)
Let $\omega\in \Omega^*_p$ be arbitrary. Suppose $\omega$ is not uniform.
Without loss generality, $\omega(L-1)\neq \omega(L)$. Let $\omega'$ be obtained from $\omega$ by replacing the values of $L-1$ and $L$ by their averages.
If $\veps_{L-1}=\veps_L$, then $f_{\omega}(\veps)=f_{\omega'}(\veps)$. Also,
\begin{equation}\label{eq:eq9}
 |f_{\omega}(\veps_1,\dotsc,\veps_{L-2},1,-1)|+
|f_{\omega}(\veps_1,\dotsc,\veps_{L-2},-1,1)|
 \geq |f_{\omega'}(\veps_1,\dotsc,\veps_{L-2},1,-1)|+
|f_{\omega'}(\veps_1,\dotsc,\veps_{L-2},-1,1)|.
\end{equation}
with equality only if $\abs{\sum_{i\leq L-2} \veps_i \omega(i)}\geq \abs{\omega(L-1)-\omega(L)}$. Indeed, denoting $a =
\sum_{i\le L-2} \veps_i \omega(i)$ and $b=\omega(L-1)-\omega(L)$ the inequality~\eqref{eq:eq9} is just
	$$ |a+b| + |a-b| \ge 2|a|\,,$$
	which is a consequence of convexity of $|\cdot|$. Furthermore, it is clear that equality holds only when $a+b$ and
	$a-b$ have the same sign, i.e. $|a| \ge |b|$.

Since $\omega\in \Omega^*_p$, it follows that the equality does hold in~\eqref{eq:eq9} for every $\veps$, 
and that $\omega'\in \Omega^*_p$ as well. From the condition for equality, we deduce that for any $\omega\in \Omega^*_p$
we have
\begin{equation}
\label{eq:omegaoptimality}
  \abs{\omega(i)-\omega(i')}\leq \min_{\veps} \abs{\sum_{j\notin\{i,i'\}} \veps_j \omega(j)}\qquad\text{for all }i\neq i'
\end{equation} \medskip

From continuity of $\omega\mapsto\tau_{\omega}$, it follows by repeated pairwise averaging, that if 
$\omega'$ is obtained from $\omega$ by replacing the values of
$\omega$ on any subset of $[L]$ by their averages, then $\omega'\in \Omega^*_p$ as well. In particular,
$\Unif\in \Omega^*_p$ and so \eqref{tauextreme} holds.\medskip

\textbf{Part \eqref{tauunique}:} Suppose that $L$ is even and \eqref{tauunique} does not hold. 
Let $\omega\in \Omega^*_p$ be non-uniform. 
Without loss of generality, 
$\omega(L-1)\neq \omega(L)$. 
Let $\omega'$ be obtained from $\omega$ by replacing values on $[L-2]$ by their averages. Since $\sum_{j\leq L-2} (-1)^{j}\omega'(j)=0$,
the measure $\omega'$ fails \eqref{eq:omegaoptimality}, and so $\omega'\notin \Omega^*_p$. Thus, $\omega \not\in
\Omega^*_p$ and hence, $\Omega^*_p = \{\Unif\}$, as claimed by \eqref{tauunique}.\medskip

\textbf{Part \eqref{taucompute}:} 
Consider a random walk on $\mathbb{Z}$ starting from $0$ that makes a step to the right with probability $p$
and to the left with probability $1-p$. Let $\Delta_{s,p}$ be the position of this walk after $s$~steps.
Relation \eqref{eq:meandiff} implies that 
\begin{equation}\label{eq:randomwalk}
\tau_{\Unif,p}=\tfrac{1}{2}-\tfrac{1}{2L}\E\abs{\Delta_{L,p}}.
\end{equation}
From \eqref{eq:randomwalk} and \cite[Eq.~(23) and~(32)]{mathworld} we obtain the formula \eqref{eq:tauL} for~$\tau_L$.\medskip

\textbf{Part \eqref{taubiased}:} In view of \eqref{tauextreme} we may restrict ourselves to the case $\omega=\Unif$. 
Because of \eqref{eq:randomwalk} and the symmetry under $p\to (1-p)$, it suffices to prove that 
\begin{equation}\label{eq:biasinduc}
\Pr[\abs{\Delta_{s,p}}\geq k]>\Pr[\abs{\Delta_{s,1/2}}\geq k] \qquad\text{ for every $p>1/2$ and every $s\geq 2$}.
\end{equation}
Since $\Delta_{s,p}$ and $\Delta_{s,1/2}$ are of the same parity as $s$, it suffices to prove \eqref{eq:biasinduc} for the case $s\equiv k\pmod 2$.
If $k=0$ or $k=1$ and $s\equiv k\pmod 2$, then both sides of \eqref{eq:biasinduc} are equal to~$1$.
The general case follows by induction on $s$ and $k$ from
\begin{align*}
  \Pr\bigl[\abs{\Delta_{s+1,p}}\geq k\bigr]&=\tfrac{1}{2}\Pr\bigl[\abs{\Delta_{s,p}}\geq k-1\bigr]+\tfrac{1}{2}\Pr\bigl[\abs{\Delta_{s,p}}\geq k+1\bigr]\\
&\qquad+(p-\tfrac{1}{2})
\bigl(\Pr[\Delta_{s,p}\in\{k,k-1\}]-\Pr[\Delta_{s,p}\in\{-k,-k+1\}]\bigr).\qedhere
\end{align*}
\end{proof}

\paragraph{Mean radii in large codes} Here we show that the average $\mrad_{\omega}(\cdot)$ over $L$-tuples in a large $C\subset \{0,1\}^n$
can be only slightly larger than $\tau_{\omega}$. In fact, we will show a generalization of this to codes of possibly small radius.
\begin{lemma}\label{lem:meancode}
Let $\omega$ be a probability measure on $[L]$.
Suppose $C\subset \{0,1\}^n$ satisfies $\rad(C)\leq p\leq \tfrac{1}{2}$. Then
\[
  \E_{w^{(1)},\dotsc,w^{(L)}\in C} \mrad_{\omega}(w^{(1)},\dotsc,w^{(L)})\leq \tau_{\Unif,p},
\]
where the expectation is over uniformly and independently chosen codewords $w^{(1)},\dotsc,w^{(L)}$ from~$C$.
\end{lemma}
\begin{proof}
Let $p_j=\Pr_{w\in C}[w_j=1]$. We have $\mrad_{\mathcal{U}[C]}(C)\le p$ from~\eqref{eq:radmrad}. Recall that one can always take $y$ attaining 
minimum in the definition of $\mrad$ to have all its coordinates in $\{0,1\}$. So, without loss of generality
(otherwise invert some coordinates), we may assume that $y$ attaining $\mrad_{\mathcal{U}[C]}(C)$ in~\eqref{eq:meanrad} is $y=0$.
Then we have $\frac{1}{n} \sum_{j\in [n]} p_j \le p$. Denote by $B(q)$ the distribution on $\{1,-1\}^L$ where each coordinate is independently
$1$ with probability $q$ and $-1$ with probability~$1-q$. Given a vector of signs $\veps=(\veps_1,\dotsc,\veps_L)\in \{1,-1\}^L$ 
define $f_{\omega}(\veps)\eqdef \sum_i \veps_i \omega(i)$.

From \eqref{eq:meandiff} and the proof of part \eqref{tauextreme} of Lemma~\ref{lem:tau} we then have 
\[
  1-2\E_{w\in C^L} \mrad_{\omega}(w)=\E_{j\in [n]} \E_{\veps\sim B(p_j)} \left\lvert f_{\omega}(\veps)  \right\rvert\geq \E_{j\in [n]} \E_{\veps\sim B(p_j)} \left\lvert f_{\Unif}(\veps)  \right\rvert
\]
By \cite[Lemma~8]{yura_better_blinovsky}, the function $p\mapsto \E_{\veps\sim B(p)} \left\lvert f_{\Unif}(\veps) \right\rvert$ is convex. Jensen's inequality
and the fact that $\tau_{\Unif,p}$ is an increasing function of $p$ on $[0,\tfrac{1}{2}]$ then complete the proof.
\end{proof}
\begin{corollary}\label{cor:meanmrad}
Let $\omega$ be a probability measure on $[L]$.
Suppose $C\subset \{0,1\}^n$ is of size $\abs{C}\geq L^2 M$ and satisfies $\rad(C)\leq p$. Then
there is an $L$-tuple $w\in C^L$ with distinct codewords such that $\mrad_{\omega}(w)\leq \tau_{\Unif,p}+1/M$.
\end{corollary}
\begin{proof}
Let $X\subset C^L$ be the set of all $L$-tuples with distinct codewords.
The corollary follows from $\Pr[w\not\in X]\leq \binom{L}{2}/\abs{C}$ and $\E_{w\in C} \mrad_{\omega}(w)\geq \Pr[w\in X]
\E_{w\in X} \mrad_{\omega}(w)$.
\end{proof}


\section{Abundance of random-like \texorpdfstring{$L$-tuples}{L-tuples}}

\begin{lemma}\label{lem:subsetbias2}
Let $\pi\colon \R^n\to \R^m$ be an orthogonal projection on a set of $m$ coordinates.
Suppose that $C\subset \{0,1\}^n$ satisfies $\rad(\pi(C))\leq \tfrac{1}{2}-\veps$. Then
there is a $C'\subset C$ of size $\abs{C'}\geq \abs{C}/2$ satisfying $\rad(C')\leq \tfrac{1}{2}-\tfrac{m}{n}\veps$.
\end{lemma}
\begin{proof} Let $\pi'$ be the projection on the remaining $n-m$ coordinates. Classify codewords $c\in C$ based on whether
$\|\pi'(c)\| \le \frac{1}{2}$ or $>\frac{1}{2}$. Without loss of generality, at least half of them (call it $C')$ satisfy
$\|\pi'(c)\|\le \frac{1}{2}$. Let $y_1 \in \R^m$ be the center attaining $\rad(\pi(C))$ and define $y\in \R^n$ to be the solution
to $\pi(y)=y_1,\, \pi'(y)=0$. We have for any $c\in C'$
\[\|y-c\|=\tfrac{m}{n} \|\pi(y)-\pi(c)\| + \tfrac{n-m}{n}
\|\pi'(y)-\pi'(c)\| \le \tfrac{m}{n}\left(\tfrac{1}{2} - \veps\right) + \tfrac{n-m}{2n} =
\tfrac{1}{2}-\tfrac{m}{n}\veps\,.\qedhere\] 
\end{proof}

For an $L$-tuple 
$x=(x^{(1)},\dotsc,x^{(L)})\in (\{0,1\}^n)^L$, we define $\type(x)\eqdef\bigl(\type(x)_u\bigr)_{u\in\{0,1\}^L}$ to be the probability distribution on $\{0,1\}^L$
with $\type(x)_u \eqdef \frac{1}{n} \#\{j: x^{(i)}_j = u_i, \forall i\in[L]\}$ (Note that $\type(x)_u = \alpha_T$ with
$T=u$ in the notation of Lemma~\ref{lem:finiteness}). The next result shows that the
only obstruction to finding a large number of $L$-tuples with approximately uniform $\type(x)$ is 
the existence of a large biased subcode.

\begin{lemma}\label{lem:uniftype} Let $L$ be fixed. For every $\veps >0$ there is a $\delta >0$ with the following property.
If $s$ is a natural number, there exist constants $M_0=M_0(s)$ and $c=c(s)$ such that for any code $C \subset \{0,1\}^n$ with $M \eqdef \abs{C} \ge M_0$ one of the following two alternatives must hold:
\begin{enumerate}
\item \label{alt:biased} $\exists C' \subset C$ such that $\abs{C'} \ge s$ and $\rad(C') \le \tfrac{1}{2}-\delta$, or
\item there exist at least $M^L - c M^{L-1}$ many $L$-tuples of distinct codewords from $C$ such that 
	\begin{equation}\label{eq:typeconst}
		|\type(x)_u - 2^{-L}| \le \veps \qquad \forall u\in \{0,1\}^L
\end{equation}	
	and, in particular
\begin{equation}\label{eq:typerad}
			|\mrad_\omega(x) - \tau_{\omega,1/2}| \le 2^{L}\veps \qquad\text{for every }\omega\in \Omega_L
\end{equation}
for each of these $L$-tuples~$x$.
\end{enumerate}
Consequently, if $C$ does not satisfy a), then the number of $L$-tuples of distinct codewords of $C$
violating~\eqref{eq:typeconst} is of size at most $c M^{L-1}$.
\end{lemma}
\begin{proof} Set $2\delta_0 \eqdef (1+2^L \veps)^{1/L} - 1$ and note that with this choice we have $|(\tfrac{1}{2} \pm
\delta_0)^L - 2^{-L}|\le \veps$. Set also $\mu \eqdef (\tfrac{1}{2}-\delta_0)^L$ and $\delta\eqdef\delta_0 \mu$.  Finally, set
$c(s) \eqdef  s 2^{L+3}$ and $M_0(s) \eqdef s 2^{L+3}$. Note that~\eqref{eq:typeconst} implies~\eqref{eq:typerad}
via~\eqref{eq:meanrad}, and so we only consider~\eqref{eq:typeconst} below.

Let us assume that a) does not hold. Then in any $C''$ with $\abs{C''} \ge 4s$ and for any orthogonal
projection $\pi_A$ on a subset of
coordinates $A \subset [n]$ there must exist a $c \in C''$ such that 
\begin{equation}\label{eq:ccd_1}
	\|\pi_A(c)\| \in (1/2-\delta_0,1/2 + \delta_0),
\end{equation}
provided that $\delta_0 \frac{|A|}{n} \ge \delta$. Indeed, if all $c \in C''$ violate~\eqref{eq:ccd_1}, then at least
half of $c\in C$ should satisfy $\|\pi_A(c)\| \le {1/2 - \delta_0}$, say. Denote this collection by $C_0$ and observe that
$\rad(\pi_A(C_0)) \le 1/2-\delta_0$ and $\abs{C_0}\ge 2s$. By
Lemma~\ref{lem:subsetbias2} there must exist $C'\subset C_0$ of size $\ge s$ such that $\rad(C') \le
\tfrac{1}{2}-\delta$, contradicting assumption.

It similarly follows that for any collection of subsets $A_1,\ldots,A_r$ with $|A_j| \ge \mu n$ for all $j\in [r]$, and any $C''$ with $\abs{C''} \ge 4sr$ there must exist $c\in C''$ such
that~\eqref{eq:ccd_1} holds simultaneously for all $A=A_j$, $j\in[r]$. Indeed, a given $c\in C''$ can violate~\eqref{eq:ccd_1} in $2r$ (two ways for each $A_j$). By the pigeonhole principle,
if all codewords in $C''$ fail~\eqref{eq:ccd_1}, then there are $2s$ that fail in the same way, and then we proceed as in the case above.

We next show that there are more than
	$$ N_1 = \prod_{j=0}^{L-1}(M-j-4s \cdot 2^j)$$
$L$-tuples $x$ of distinct codewords from $C$ that satisfy~\eqref{eq:typeconst}. Indeed, at least
$M-4s$ codewords $x^{(1)}$ have $|\|x^{(1)}\|-\tfrac{1}{2}| \le \delta_0$. Once one such codeword $x^{(1)}$ is selected,
let $A_0=\{j\in [n]: x^{(1)}_j = 0\}$ and $A_1 = A_0^c$. Each of these two subsets has cardinality $\ge
n(\tfrac{1}{2}-\delta_0) \ge \mu n$. By the argument above, there are
more than $M-1-4s\cdot 2$ codewords $x^{(2)}$ not equal to $x^{(1)}$ such that projections of $x^{(2)}$ on $A_0$ and $A_1$
both have weights $\in [\tfrac{1}{2}-\delta_0, \tfrac{1}{2}+\delta_0]$. Selecting one such $x^{(2)}$, we define
partitions $A_0 = A_{00} \cup A_{01}$ and $A_1=A_{10}\cup A_{11}$ 
according to the values of coordinates of $x^{(1)}$ and
$x^{(2)}$. Proceeding similarly, we construct $x^{(3)},\ldots,x^{(L)}$. The resulting $L$-tuple has distinct elements and satisfies
$$ (\tfrac{1}{2}-\delta_0)^L \le \type(x)_u \le (\tfrac{1}{2} +\delta_0)^L \qquad \forall u\in\{0,1\}^L\,,$$
which by the choice of $\delta_0$ implies that it satisfies~\eqref{eq:typeconst} as well.

Note that for $M\geq \max_j k_j$ we have
\[ \prod_{j=0}^{L-1} (M-k_j) = M^L\prod_{j=0}^{L-1}(1-k_j/M) \geq M^{L} - M^{L-1} \sum_{j=0}^{L-1} k_j\,.\]
Setting $k_j = s 2^{j+3} \ge j+4s\cdot 2^j$ we
obtain 
$$ N_1 \ge M^L - c M^{L-1}\,,$$
provided $M\ge M_0$, completing the proof of the first part.

The final statement of the lemma follows from the fact that there are at least $M^L-cM^{L-1}$ many $L$-tuples
satisfying~\eqref{eq:typeconst}.
\end{proof}

\section{Proof of Theorem~\protect\ref{thm:even}}
Let $L$ be even, and suppose $C\subset \{0,1\}^n$ is an \Ldecodable/ code of radius~$\tau_L+\veps$.  We wish to prove
that $\abs{C}=O(\veps^{-1})$. Let $\rho_L(C) = \min_{x\in C^L} \rad(x)$ 
with the minimum taken over all $L$-tuples $x$ with
distinct elements.
Unlike $\rho_L(C)$, the $L$-radius of a code (denoted $\tilde \rho_L(C)$) is not a well-behaved quantity. Sadly, our assumptions
on $C$ do not imply that $\rho_L(C)\geq \tau_L$. For example, if $L=2$ then the radius of
$\{000,100\}$ is $1/3>1/4=\tau_2$ whereas 
$\rad(000,100)=1/6$. To get around this, we use the pigeonhole principle
to find a subcode $C'$ of size $\abs{C'}\geq 2^{-8L}\abs{C}$ consisting of codewords with the same prefix of length~$8L$. Removing
the common prefix yields a code of block length $n-8L$ whose $L$-radius is at least 
\[
  \frac{n}{n-8L}(\tau_L+\veps)\geq (1+8L/n)\tau_L+\veps\geq \tau_L+\veps+2L/n.
\]
By Proposition~\ref{prop:rounding} we have $\rad(x)\geq \tau_L+\veps$ for every $L$-tuple $x$ of distinct codewords from this
new code. With slight abuse of notation, we rename this new code~$C$ (and adjust the value of $n$ accordingly).

\begin{lemma}\label{lem:smallrad} 
Let $C'$ be any code with $\rho_L(C') \ge \tau_L$. If $\rad(C')\leq \tfrac{1}{2}-\delta$ then $\abs{C'}<s$ for some
$s$ depending on~$\delta$.
\end{lemma}
\begin{proof} Identify $L$-element subsets of $C'$ with ordered $L$-tuples by fixing some (arbitrary) ordering on~$L'$. 
Elements of such $L$-tuples are distinct. For every $x\in (C')^L$, which is an $L$-tuple with distinct elements,
there is $\omega\in \Omega_L'$ that solves~\eqref{eq:radfin}. This gives a coloring of $L$-element subsets of $C'$ into
$\abs{\Omega_L'}$ colors. From finiteness of $\Omega'_L$ and the hypergraph version of Ramsey's theorem \cite[Theorem~2]{ramseybook},
it follows that, if $C'$ is large enough, then there is a monochromatic subset $C''\subset C'$ of size exceeding $\frac{L^2}{\tau_L-\tau_{\Unif,p}}$,
where $p\eqdef\tfrac{1}{2}-\delta$.

Let $\omega\in \Omega'_l$ be the color of $C''$, i.e., $\mrad_\omega(x) \ge \tau_L $ for any $L$-tuple $x\in (C'')^L$ with distinct elements.
Since $\rad(C'')\leq \rad(C')\leq p$, it follows from Corollary~\ref{cor:meanmrad} and the bound
$\tau_{\Unif,p}<\tau_L$ of Lemma~\ref{lem:tau} that $\abs{C''} \leq \frac{L^2}{\tau_L-\tau_{\Unif,p}}$. 
The contradiction shows that $C'$ cannot be arbitrarily large.
\end{proof}

Let $H$ be the set of all $L$-tuples $x\in C^L$ such that
$\mrad_{\omega}(x)> \tau_L$ for some $\omega\neq \Unif$.
\begin{lemma}
We have $\abs{H}\leq c_L \abs{C}^{L-1}$ for some constant $c_L$ that depends only on~$L$. 
\end{lemma}
\begin{proof}
Let $\veps=2^{-L}\min \{\tau_L-\tau_{\omega,1/2} : \omega\in \Omega_L',\ \omega\neq\Unif\}$. Since $\Omega_L'$ is finite,
part \eqref{tauunique} of Lemma~\ref{lem:tau} implies that $\veps>0$.
So, let $\delta$ be as in Lemma~\ref{lem:uniftype} for this value of~$\veps$. Let $s$ be the bound
from Lemma~\ref{lem:smallrad}. Note that by the choice of $\veps$, the set $H$ consists entirely of the $L$-tuples
violating~\eqref{eq:typerad} and hence~\eqref{eq:typeconst}. By the choice of $s$, alternative \eqref{alt:biased} in
Lemma~\ref{lem:uniftype} is impossible. Therefore, by the last statement of the latter Lemma, we have 
$\abs{H}\le c(s) \abs{C}^{L-1}$.
\end{proof}

\begin{proof}[Proof of Theorem~\ref{thm:even}]
Call an $L$-tuple $x\in C^L$ \emph{good} if all of its codewords are distinct, and $x\notin H$. 
For a randomly chosen $L$-tuple $x\in C^L$, the probability that $x^{(i)}=x^{(i')}$ for some $i\neq i'$
is less than $L^2/\abs{C}$. By the preceding lemma and finiteness of the set $\Omega_L'$, the probability 
that $x\in  H$ is also $O(1/\abs{C})$. So a random $x$
is good with probability $1-O(1/\abs{C})$. Lemma~\ref{lem:meancode} then implies that 
\[
  \Pr[x\text{ is good}] \E_{\text{good }x} \mrad_{\Unif}(x)\leq \tau_{\Unif,1/2}=\tau_L\,.
\]
On the other hand, for a good $L$-tuple we have $\rad(x)=\mrad_{\Unif}(x)$ and thus the expectation is lower bounded by
$\tau_L+\veps$. In all, we conclude that $\frac{\veps}{\tau_L+\veps}=O(1/\abs{C})$, completing the proof.
\end{proof}

\section{Proof of Theorem~\ref{thm:simpleconstruction}}
\begin{proof}[Proof of Theorem~\ref{thm:simpleconstruction}]
Recall that $M=\binom{2m}{m}$. 
Consider an $2m$-by-$M$ matrix with $\{0,1\}$ entries 
whose columns are all possible vectors consisting of exactly $m$ ones. The $2m$ rows of the matrix are the codewords of a code $C\subset \{0,1\}^M$.
We claim that $\mrad_{\Unif}(x)\geq \tau_L+c_L/2m+O(m^{-2})$ for every $L$-tuple $x$ of distinct codewords from~$C$.

By symmetry, $\mrad_{\Unif}(x)$ is independent of the 
actual choice of $x$. So, fix any $x$, and pick $j$ at random from $[M]$. 
Let $0_j$ be the number of these codewords that have $0$ in the $j$'th column. Similarly,
let $1_j$ be the number of these codewords that have $1$ in the $j$'th column. Let $X_j=\min(0_j,1_j)/L$. 
Note that $\mrad_{\Unif}(x)=\E X_j$ by \eqref{eq:meanrad}.

Suppose $L=2k+1$ is odd. Then 
\begin{align*}
  \E_j X_j&=\frac{1}{2k+1}\sum_{l\leq k}l\cdot \Pr[\min(0_i,1_i)=l] \\&= 
  \binom{2m}{m}^{-1}\frac{1}{2k+1}\sum_{l\leq k} 2l\binom{2k+1}{l}\binom{2m-2k-1}{m-l} 
\\&=\binom{2m}{m}^{-1}\sum_{1 \leq l\leq k} 2\binom{2k}{l-1}\binom{2m-2k-1}{m-l}\\
 &=\sum_{1 \leq l\leq k} 2\binom{2k}{l-1}\frac{m(m-1)\dotsb(m-l+1)\cdot m(m-1)\dotsb(m-2k+l)}{2m(2m-1)\dotsb(2m-2k)}\\
\intertext{which, as $m\to\infty$, is }
 &=\sum_{1 \leq l\leq k} \binom{2k}{l-1}2^{-2k}\left[1+\frac{1}{2m}\binom{2k+1}{2}-\frac{1}{m}\binom{l}{2}-\frac{1}{m}\binom{2k-l+1}{2}+O(m^{-2})\right]\\
 &=\tau_{2k+1}+2^{-2k-1}k\binom{2k}{k}/2m+O(m^{-2})
\end{align*}
In the last equality here we used the expression for $\tau_{2k+1}$,
the formula for the variance of the binomial random variable
$B(2k,1/2)$, and the known expression for the expected distance of
a balanced random walk of $2k$ steps from the origin.

Similar computations hold if $L=2k$. Denote by $\sum\nolimits'$ the sum in which the last summand is halved. The expected value of $X_j$ is
\begin{align*}
  \binom{2m}{m}^{-1}\frac{1}{2k}&\sideset{}{'}\sum_{l\leq k} 2l\binom{2k}{l}\binom{2m-2k}{m-l} \\
&=\binom{2m}{m}^{-1}\sideset{}{'}\sum_{l\leq k} 2\binom{2k-1}{l-1}\binom{2m-2k}{m-l}\\
&=\sum_{l\leq k}2\binom{2k-1}{l-1}\frac{m(m-1)\dotsb(m-l+1)\cdot m(m-1)\dotsb(m-2k+l+1)}{(2m)(2m-1)\dotsb(2m-2k+1)}\\
&=\sum_{l\leq k}\binom{2k-1}{l-1}2^{-2k+1}\left(1+\frac{1}{2m}\binom{2k}{2}-\frac{1}{m}\binom{l}{2}-\binom{2k-l}{2}+O(m^{-2})\right)\\
&=\tau_{2k}+2^{-2k}k\binom{2k-1}{k}/2m+O(m^{-2}).\qedhere
\end{align*}
\end{proof}

\section{Proof of Theorem~\ref{thm:lthree}}\label{sec:proof_three}

We start with the proof of the upper bound, following the approach
of Konyagin in \cite{Kon}. Let $C$ be \threedecodable/ code of vectors in $\{0,1\}^n$ of radius at most
$\tau_3+\veps=1/4+\veps$. By Proposition~\ref{prop:triangles} this implies
that among any $3$ codewords in $C$ there are two of distance
at least $(1/2+2\veps)n$. For each codeword $x=(x_1,x_2, \ldots
,x_n)$ define a vector $v=(v_1,v_2, \ldots ,v_n)$ in
the Euclidean space $R^n$ by $v_i=\frac{(-1)^{x_i}}{\sqrt n}$.
Note that each such vector is of unit norm, and among any 
three vectors there are two whose inner product is at most $-4 \veps$.
Let $V$ be the set of all the vectors obtained from the words in
$C$ and put $|V|=m$. Our objective is to show that
$m \leq O(1/\veps^{3/2})$. Let $H=(V,E)$ be the graph whose set of
vertices is $V$ in which two vertices $u,v$ are connected iff their
inner product is larger than $-4 \veps$.  Fix a vertex $v \in V$
and let $W=N(v)$ be the set of all its neighbors in $H$. Note that
the inner product between any two vertices in $W$ is at most
$-4 \veps$. Therefore, if $d=|W|$ is the degree of $v$ in $H$ and
$\|v\|$ denotes the Euclidean $2$-norm of a vector $v$, then
\begin{equation}\label{eq:konyagin0}
0 \leq \biggl\lVert\sum_{u \in W} u \biggr\rVert^2 \leq d -d(d-1) 4 \veps
\end{equation}
implying that  $d \leq \frac{1}{4\veps}+1$ and also implying that
\[
\biggl\lVert\sum_{u \in W} u \biggr \rVert^2 \leq d -d(d-1) 4 \veps 
=\frac{1}{4 \veps} (4\veps d)(1+4\veps-4 \veps d)
\leq \frac{(1+4\veps)^2}{16 \veps}.
\]
Therefore, by Cauchy--Schwarz, for every $v \in V$
\begin{equation}\label{eq:konyagin1}
\sum_{u \in N(v)} \langle v,u \rangle \leq \biggl\lVert\sum_{u \in N(v)} u \biggr\rVert
\leq \frac{1+4\veps}{4 \sqrt {\veps}}.
\end{equation}
This gives the following (which can be slightly improved, but as
this only changes the error term we prefer to present the simple
version below):
\begin{align*}
0 \leq \biggl\lVert\sum_{v \in V} v \biggr\rVert^2&=m +\sum_{v \in V} \sum_{u \in N(v)}
\langle v, u \rangle + \sum_{u \neq v \in V, uv \not \in E}
\langle v,u \rangle\\
&\leq m+m\frac{1+4\veps}{4 \sqrt {\veps}}-m\left(m-\frac{1}{4\veps}-2\right) 4
\veps.
\end{align*}
By the last inequality
\[
\left(m-\frac{1}{4\veps}-2\right) 4 \veps \leq 1+\frac{1+4\veps}{4 \sqrt
{\veps}},
\]
implying that
\begin{equation}\label{eq:konyagin2}
m \leq \frac{1}{16 \veps^{3/2}}+O\left(\frac{1}{\veps}\right).
\end{equation}
This completes the proof of the upper bound.

\vspace{0.2cm}

\noindent
We proceed with the proof of the lower bound by describing an
appropriate construction.
Let $G = (V,E)$ be a graph on $m$ vertices, suppose it is a Cayley graph of an elementary abelian
$2$-group $\Z_2^r$, let $A$ be its adjacency matrix, and let $d = \lambda_1\geq  \lambda_2\geq \dotsb \lambda_m= -\lambda$ be its eigenvalues,
where $d$ is the degree of regularity and~$-\lambda$ is the smallest eigenvalue. Assume, further, that 
$G$ is triangle-free. As $G$ is a Cayley graph of an elementary abelian $2$-group, it has an orthonormal
basis of eigenvectors $v_1,v_2,...,v_m$ in which each coordinate of each vector is in $\{-1/\sqrt{m},1/\sqrt{m}\}$.
Define $B = (A+\lambda I)/\lambda$ where $I$ is the $m$-by-$m$ identity matrix. Then $B$ is a positive semidefinite
matrix, its diagonal is the all-$1$ vector, its eigenvalues are $\mu_i= (\lambda_i+\lambda)/\lambda$ and the corresponding
eigenvectors are the vectors~$v_i$. Let $P$ be the $m$-by-$m$ orthogonal matrix whose columns are the
vectors $v_i$, and note that the first $v_1$ is the constant vector~$1/\sqrt{m}$. Let $D$ be the diagonal matrix
whose diagonal entries are the eigenvalues~$\mu_i$ and let
$\sqrt{D}$ denote the diagonal matrix whose entries
are~$\sqrt{\mu_i}$. Then 
$P^tBP=D$ and thus $B = (P\sqrt{D})(\sqrt{D}P^t)$.

The rows of the matrix $P\sqrt{D}$ are row-vectors $x_1,x_2,\dotsc,x_m$ where $x_i=(x_{i1},x_{i2},\dotsc,x_{im})$.
Note that for each $j$, $x_{ij}\in \{-\sqrt{\mu_j/m},\sqrt{\mu_j/m}\}$ for all $i$, and that $x_{i1}$ is positive for all~$i$. 
In addition $x_i x_j^t = B_{ij}$ for all $i,j$ meaning that the $\ell_2$-norm of each vector $x_i$ is $1$ and that among any three
vectors $x_i$ there is an orthogonal pair. Let $y_i$ be the vector obtained from $x_i$ by removing its
first coordinate (the one which is $\sqrt{\mu_1/m} 
=\sqrt{(d + \lambda)/m\lambda}$). Then each $y_i$
is a vector of $\ell_2$-norm $1 - \mu_1/m$ and among any three of them there is a pair with inner product $-\mu_1/m$. We can
normalize the vectors by dividing each entry by
$\sqrt{1 - \mu_1/m}$ to get $m$ unit vectors $z_1,z_2,\dotsc,z_m$,
where any three of them contain a pair with inner product $-\delta$, where $\delta = \mu_1/(m-\mu_1)$. Moreover,
for the vectors $z_i= (z_{ij})$, for each fixed $j$ the absolute value of all $z_{ij}$ is the same for all~$i$, denote
this common value by~$t_j$. We can now use the vectors $z_i$ to define functions mapping $[0,1]$ to
$\{1,-1\}$ as follows. Split $[0,1]$ into disjoint intervals $I_j$ of length $t_j^2$
and define $f_i$ to be $\sign(z_{ij})$ on the interval~$I_j$. It is clear that the $\ell_2$-norm of each $f_i$ is $1$ and the inner product between $f_i$
and $f_j$ is exactly that between $z_i$ and~$z_j$. In particular, each three functions $f_i$ contain a pair
whose inner product is at most~$-\delta$.

One can replace the functions by vectors of $1,-1$ with essentially the same property, using
an obvious rational approximation to the lengths of the intervals.

The graph in \cite{noga_trianglefree} is a triangle-free Cayley graph of an elementary abelian $2$-group with
$d = \bigl(1/4+o(1)\bigr)m^{2/3}$ and $\lambda = \bigl(9+o(1)\bigr)m^{1/3}$. This gives us $\delta = (1/36+o(1))m^{-2/3}$ and hence the
number of vectors is $m = \Theta\bigl((1/\delta)^{3/2}\bigr)$. Setting $\delta = 4 \veps$ this gives a binary code with $m = \Theta\bigl((1/\veps)^{3/2}\bigr)$ codewords
of length $n$ so that among any three codewords there are two 
such that the Hamming distance between them 
is at least $(1/2 + 2\veps)n$. According to
Proposition~\ref{prop:triangles},
this means that the code is \threedecodable/ with $\tau = {1\over4} + \veps$, and thus $\maxcode_{\veps}(L=2) =
\Omega(\veps^{-{3\over2}})$.

\section{Spherical codes in the Hilbert space}\label{sec:hilbert}
Let us now consider a similar question for the case of the real Hilbert space $\Esp$ (the space of  square-summable
sequences of real numbers). Similar to the binary alphabet, we may motivate the question by the desire to construct
a maximal number $M$ of unit-energy signals, such that when one of them is sent and adversarial noise of bounded energy is
applied, it is still possible to reconstruct the original signal, to within a list of size $<L$. We also note that
results on adversarial-noise lead to bounds for the average-noise variation, as propounded in~\cite[Section XII]{CS59}.
We proceed to formal definitions.

We shall employ the same notation as in the rest of the paper, but with the meaning adapted
to spherical codes. For example, we denote the norm in $\Esp$ by $\|\cdot\|$.  We redefine $\rad(x)$ similarly: 
for an arbitrary $L$-tuple $x=(x^{(1)},\dotsc,x^{(L)})\in \Esp^L$ we define
$$ \rad(x) = \min_{y\in \Esp} \max_j \|x^{(j)}-y\|\,, \quad \diam(x) = \max_{i,j} \|x^{(i)} - x^{(j)}\|\,.$$
Recall Jung's theorem~\cite[(2.6)]{DGK63}: For any $L$-tuple $x$ we have
\begin{equation}\label{eq:jung}
	\rad(x) \le \sqrt{L-1 \over 2L} \diam(x) 
\end{equation}
with equality if and only if $x^{(1)},\dotsc,x^{(L)}$ are the vertices of an $(L-1)$-simplex, i.e., when $x$ consists of $L$ vectors 
with pairwise distances all equal.

A \emph{spherical code} $C$ is a finite collection of unit-norm
vectors in $\Esp$ and its $L$-radius $\rho_L(C)$ is the minimum
value of 
$\rad(x)$ among all $L$-tuples $x$ of distinct elements of $C$. We define 
\[
  \maxcode_L(\veps)\eqdef \sup \bigl\{ \abs{C} : \rho_L(C) \ge \tau_L + \veps\bigr\}\,,
\]
where in this section $\tau_L \eqdef \sqrt{L-1\over L}$. Our formulation corresponds to the problem of packing balls
$B(x,r) \eqdef \{y \in \Esp: \|x-y\|\le r\}$ centered on the unit sphere so that no point of $\Esp$ is covered by more
than $(L-1)$ of them. Another equivalent way is to consider the problem of packing spherical caps $C(x,\alpha)=\{y: \|y\|=1,
\mla y, x \mra \ge \cos\alpha\}$, where $\|x\|=1$,  with the requirement that no point \textit{of the unit sphere} is covered by more than
$(L-1)$ of them.

A classical result of Rankin~\cite{rankin1955closest} solves
the case $L=2$:
\begin{equation}\label{eq:rankin}
	\maxcode_2(\veps) = \left\lceil 1+{1\over 2\sqrt{2} \veps + 2\veps^2} \right \rceil = \Theta \left({1\over
\veps}\right).
\end{equation}

For $L>2$, Blachman and Few~\cite{blachman1963multiple} proved that if $\Esp$ is replaced by $\mathbb{R}^n$ then codes
with $\rho_L(C) > \tau_L$ have size polynomial in $n$, while for $\rho_L(C) < \tau_L$ exponentially large codes exist.
This was improved by Blinovsky~\cite{blinovsky1999multiple}, who demonstrated that codes with 
$\rho_L(C) > \tau_L + \veps$, $\veps > 0$  have a finite upper bound on their 
size independent of $n$. His proof relied on 
the Ramsey theorem and can be condensed as follows:

\begin{proposition}[\cite{blinovsky1999multiple}] For any $\veps > 0$ $\maxcode_L(\veps)$ is finite.
\end{proposition}
\begin{proof} Consider a code $C$ with $\rho_L(C) \ge \tau_L + \veps$. 
Fix an integer $q \ge 1$, and break $[0,2]$ into $q$ intervals 
of size $2\over q$. Consider a code $C$
and label each pair $(c,c') \in {C\choose 2}$ according to the interval 
which contains $\|c-c'\|$. 
By Ramsey's theorem if $C$ is sufficiently large then there should exist a large subcode $C'$ whose all pairwise
distances are in $[a, a+\tfrac{2}{q})$. From~\eqref{eq:rankin} we have $a \le \sqrt{2} + O(1/|C'|)$ and
from~\eqref{eq:jung} we get that $\rho_L(C) \le \rho_L(C') \le \tau_L + O(1/|C'|) + O(1/q)$ and hence $|C'| \le
O(1/\veps)$ when $q = O(1/\veps)$.
Consequently, $C$ cannot be arbitrary large for a given $\veps>0$.
\end{proof}

Our main result on spherical codes is the following.
\begin{theorem}\label{th:hilb} For any $L\ge 2$ there exist constants $c_1,c_2 > 0$ such that for all $\veps > 0$
\begin{equation}\label{eq:h1}
		c_1 \veps^{-1} \le \maxcode_L(\veps) \le c_2 \veps^{-{L^2 - L + 2\over 2L}}\,.
\end{equation}	
\end{theorem}

Before proving the theorem we establish two auxiliary lemmas.
\begin{definition}
Call a collection $S$ of unit vectors an \emph{$(m,\epsilon)$-system} if among any
$m$ distinct elements $x_1,\ldots,x_m \in S$ there exists a
pair with $\mla x_i, x_j\mra <-\epsilon$.
\end{definition}
\begin{lemma}\label{lem:konyagin} For each $m$ there exists $C_m>0$, such that the size of any $(m,\epsilon)$-system
$S$ is at most $C_m\epsilon^{-{m\over 2}}$
and 
	$$ \|\sum_{x\in S} x\| \le C_m \epsilon^{-{m-1\over 2}}\,.$$
\end{lemma}
\begin{proof} For $m=2$ this follows from~\eqref{eq:rankin} and~\eqref{eq:konyagin0}. For $m=3$ this was shown above
in~\eqref{eq:konyagin1} and~\eqref{eq:konyagin2}, essentially by reducing to $m=2$. In general, for arbitrary $m$ we can
define a graph with vertices $S$ as in the proof of~\eqref{eq:konyagin2} and notice that the neighborhood $\mathcal{N}(v)$ is an
$(m-1,\epsilon)$-system and then apply induction.
\end{proof}

\begin{lemma}\label{lem:hil_rad} For any $L\ge 3$ there exists a non-negative function $\tau(\gamma) = {2\gamma\over L^2 -L-2} +
O_L(\gamma^2)$, $\gamma \in [0,1]$, with the following property. Consider 
any $L$-tuple $x=(x_1,\ldots,x_L)$ of unit-norm vectors with $\rad(x) \ge \tau_L$. If $\mla
x_1,x_2 \mra \ge \gamma\ge 0$
then there exist $i,j$ such that $\mla x_i,x_j \mra \le -\tau(\gamma)$.
\end{lemma}
\begin{proof} Entirely like in~\eqref{eq:radmrad} we can prove
\begin{equation}\label{eq:rad2_minmax}
\begin{aligned}
	\rad(x)^2 &= \max_{\omega \in \Omega_L} \min_{y\in\Esp} \E_{i\sim \omega} \|x_i - y\|^2 = \max_{\omega \in \Omega_L} \min_{y\in\Esp} \left( 1 - 2\Bigl\langle \sum_i \omega_i x_i,y \Bigr\rangle + \|y\|^2 \right)  \\ &=\max_{\omega} \left(1-\Bigl\|\sum_{i}
	\omega_i x_i\Bigr\|^2\right) = 1-\min_{\omega \in \Omega_L} V(\omega)\,,
\end{aligned}
\end{equation}
where $V(\omega)=\sum_{i,j} v_{i,j} \omega_i \omega_j$ is the quadratic form corresponding to the
Gram matrix of $x$ with $v_{i,j}=\mla x_i,x_j \mra$. 

Fix some $0\le\tau\le{1\over L-1}$ and suppose now that $x$ is such that $\langle x_i,x_j\rangle \ge -\tau$ for all $i,j$. We will 
show that for some function $\tau(\gamma)$ if $\tau < \tau(\gamma)$ then $\rad(x) < \tau_L$. To that end, we introduce another quadratic form $U(\omega) = \sum_{i,j} u_{i,j} \omega_i \omega_j$ with 
\begin{equation}\label{eq:uquad}
	u_{i,j} = \begin{cases} 1, & i=j,\\
			 \gamma, & \{i,j\} =\{1,2\},\\
			 -\tau, & \text{otherwise}.
		\end{cases}
\end{equation}		
Note that according to assumptions $v_{i,j} \ge u_{i,j}$ and, therefore, on $\Omega_L$ we have $V(\omega) \ge
U(\omega)$, and 
$$ \min_{\Omega_L} V(\omega) \ge \min_{\Omega_L} U(\omega)\,.$$
We next show that $U$ is non-negative definite for all $0\le \tau \le {1\over L-1}$ and all $-{1\over L-1}\le \gamma \le
1$. From convexity of the PSD cone, it
is sufficient to check the four corners. For $\tau=0$ the statement is clear. For
$\tau = {1\over L-1}$ we consider the two endpoints: $\gamma = -{1\over L-1}$, $\gamma=1$. For
$\gamma = -{1\over L-1}$ the resulting quadratic form equals $U_1(\omega) = \sum_{i} \omega_i^2 - {1\over L-1} \sum_{i\neq j}
\omega_i \omega_j$ and 
corresponds to the Gram matrix of unit-norm vectors forming an $(L-1)$-simplex centered at the origin. Consequently, $U_1$ is positive definite. Similarly,
for $\gamma=1$, the quadratic form corresponds to Gram matrix of the following collection: take unit-norm vectors
forming an $(L-1)$-simplex, delete one vector and add a copy of another. The resulting quadratic form is non-negative
definite.

Since $U$ is convex, we could evaluate
$\min_\omega U(\omega)$ by arguing that optimal assignment is symmetric (has equal coordinates $3,\ldots,n$ and $1,2$).
Instead we prefer to proceed indirectly and show another useful property of radii in Hilbert space.


Since $U\succeq 0$, it is a Gram matrix of some other $L$-tuple $x'$ of unit-norm vectors and we know 
\begin{equation}\label{eq:radp_rad}
	\rad(x')\ge \rad(x)\,.  
\end{equation}
We temporarily forget about the special form of $U$, as defined in~\eqref{eq:uquad}, and view it as a generic Gram
matrix of \textit{some} $L$-tuple $x'$ of unit-norm vectors with the property that $|\mla x'_i,
x'_j\mra| \le \theta$ for $i\neq j$. We will prove
\begin{equation}\label{eq:radmrad_hil}
	\rad(x')^2 = \tau_L^2 - {1\over L^2} \sum_{i\neq j} \mla x'_i, x'_j \mra + O(\theta^2)\,,
\end{equation}
where the $O(\cdot)$ term is uniform in $x'$. Note that the first two terms of the expression in~\eqref{eq:radmrad_hil} 
correspond to $\omega=\Unif$ in~\eqref{eq:rad2_minmax}. As $\theta \to 0$ the $L$-tuple $x'$ becomes very
close to $L$ orthogonal vectors, and hence in~\eqref{eq:rad2_minmax} we expect that the optimal $\omega=\Unif +
O(\theta)$, cf.~\eqref{eq:rh_om}. Since we are operating 
near the minimum of the quadratic form, the $O(\theta)$ deviation
of $\omega$ translates to $O(\theta^2)$ deviation for the value of $U$.

Proceeding to a formal proof of~\eqref{eq:radmrad_hil}, first notice that if $\omega_1=0$ then as $\theta\to 0$ we must have $1-U(\omega)\le
\tau_{L-1} + o(1)$ (since we are considering only $L-1$ almost orthogonal vectors). But $1-\min_\omega U(\omega)$ 
tends to $\tau_L > \tau_{L-1}$, implying that for 
all sufficiently small $\theta$, the minimizer of $U(\omega)$ is in the
interior of $\Omega_L$. At the optimal point $\omega^*$ the gradient of $U$ is proportional to 
a vector of all ones $\ones$, from where we find 
\begin{equation}\label{eq:rh_om}
	\omega^* = c (I_L+\Delta)^{-1} \ones\,,
\end{equation}
where $(I_L+\Delta)$ is the matrix of $U$, and the 
normalizing constant $c$ is found from $\mla \omega^*, \ones \mra =
1$ yielding $c=\mla (I_L+\Delta)^{-1} \ones, \ones \mra^{-1}$. Altogether, we get 
$$ U(\omega^*) = \mla (I_L + \Delta) \omega^*, \omega^* \mra = c = \mla I_L \ones, \ones \mra +  \mla \Delta \ones,
\ones\mra +
O(\theta^2)\,.$$
Finally, since $\rad(x')^2 = 1-U(\omega^*)$ we get~\eqref{eq:radmrad_hil}.

To complete the proof of the Lemma, note that from~\eqref{eq:radp_rad},~\eqref{eq:radmrad_hil} and~\eqref{eq:uquad} we have
$$ \rad(x)^2 \le \tau_L^2 - {1\over L^2}\bigl(2\gamma - (L^2-L-2)\tau\bigr) + O(\gamma^2) + O(\tau^2)\,.$$
Consequently, for appropriately defined $\tau(\gamma)$, if $\tau < \tau(\gamma)$ we should have $\rad(x) < \tau_L$.
Furthermore, as $\gamma\to0$ we have that $\tau(\gamma) = {2\gamma\over L^2 -L-2} +
O_L(\gamma^2)$, as claimed.
\end{proof}

\begin{proof}[Proof of Theorem~\ref{th:hilb}] Consider a regular $(M-1)$-simplex of unit vectors in $\Esp$. The pairwise distances are equal $\sqrt{2M\over
M-1}$ and thus from~\eqref{eq:jung} we have that the radius of any $L$-tuple is at least $\tau_L \sqrt{M\over M-1} = \tau_L
+ \Omega(1/M)$, proving the lower bound in~\eqref{eq:h1}.\smallskip

We proceed to the upper bound. Fix a code $C$ with $\rho_L(C)\ge \tau_L + \veps$.  The main idea is again essentially due to Konyagin: fixing one point $c\in C$ and considering its
close neighbors, we notice that the radius constraint (cf.\ Lemma~\ref{lem:hil_rad}) introduces repulsion between these neighbors (that is they should
be widely separated among themselves) and consequently, neighborhoods can not be too large. 

We proceed with the argument. First, by~\eqref{eq:jung} any
 $L$-tuple with $\rad(x) \ge \tau_L + \veps$ also satisfies $\diam(x) \ge \sqrt{2} + \sqrt{2\over \tau_L} \veps$, and
thus the code $C$ is also an $(L, \epsilon')$-system, with $\epsilon' =  {2\over \sqrt{\tau_L}} \veps$.

Next, let $\epsilon_1 = \veps^{\frac{L-1}{L}}$ and $\epsilon_2 = \tau(\epsilon_1)$, where $\tau(\cdot)$ is
from Lemma~\ref{lem:hil_rad}. We consider two types of neighbors $c$ of each point $c_i \in C$, depending on 
\begin{equation}\label{eq:ell2_s0}
	-\epsilon' \le \langle c,c_i\rangle \le \epsilon_1, \mbox{~or~} \langle c,c_i\rangle > \epsilon_1\,.
\end{equation}
Let $\mathcal{N}'(c_i)$ and $\mathcal{N}''(c_i)$ be the two respective sets of neighbors. The rest of the points are ``far away'' from $c_i$
and satisfy
	\begin{equation}\label{eq:ell2_s0a}
		\langle c,c_i\rangle < - \epsilon'\,.
\end{equation}	
First, notice that since $C$ is an $(L,\epsilon')$-system, we have that $\mathcal{N}'(c_i)\cup \mathcal{N}''(c_i)$ is an
$(m,\epsilon')$-system with $m\eqdef L-1$. Thus from Lemma~\ref{lem:konyagin}
	\begin{equation}\label{eq:ell2_s1}
		|\mathcal{N'}(c_i)| \le |\mathcal{N}'(c_i) \cup \mathcal{N}''(c_i)| \le C_m\epsilon'^{-{m\over 2}}\,.
\end{equation}	
Second, take any $m=(L-1)$ distinct points in $\mathcal{N}''(c_i)$. Adding $c_i$ to this $m$-tuple and applying
Lemma~\ref{lem:hil_rad} to the resulting $L$-tuple, we conclude that $\mathcal{N}''(c_i)$ is an
$(m,\epsilon_2)$-system. Therefore, from Lemma~\ref{lem:konyagin} we have
	\begin{equation}\label{eq:ell2_s2}
		\Bigl\|\sum_{c\in \mathcal{N}''(c_i)} c \Bigr\| \le C_m \epsilon_2^{-{m-1\over 2}}\,.
\end{equation}	
Consider
	\begin{align} \Bigl\mla c_i, \sum_{c\in C} c \Bigr\mra &= 1 + 
		\Bigl\mla c_i, \sum_{c\in\mathcal{N}''(c_i)}c \Bigr\mra + \Bigl\mla c_i, \sum_{c\in \mathcal{N}'(c_i)} c \Bigr\mra +
				 \Bigl\mla c_i, \sum_{c\not\in \mathcal{N}'\cup \mathcal{N}'' \cup \{c_i\}}c \Bigr\mra \\
				&\le 1 + C_m \epsilon_2^{-{m-1\over 2}} + C_m \epsilon_1 \epsilon'^{-{m\over2}} -
				\epsilon' (|C| - 1-C_m \epsilon'^{-{m\over2}})\label{eq:ell2_s3}
\end{align}
	where the second term is estimated by 
Cauchy--Schwarz and~\eqref{eq:ell2_s2}, the third term is by 
the definition of
	$\mathcal{N}'(c_i)$ and~\eqref{eq:ell2_s1}, and the fourth term is the combination of~\eqref{eq:ell2_s0a} and 
	the bound in~\eqref{eq:ell2_s1}.

Summing~\eqref{eq:ell2_s3} over all $c_i \in C$ and using $\sum_{c_i, c\in C} \mla c_i, c\mra \ge 0$ we get 
$$ \epsilon' |C| \le 1+\epsilon' + C_m \epsilon_2^{-{m-1\over 2}} + C_m \epsilon_1 \epsilon'^{-{m\over2}} +
	C_m \epsilon'^{1-{m\over2}}\,,$$
from where, recalling that $\epsilon_1 \asymp \epsilon_2 \asymp \veps^{L-1\over L}$ and $\epsilon' \asymp \veps$ we get
that the first two terms and the last are negligible compared to the third and fourth, which are comparable and $\asymp
\veps^{-{(L-1)(L-2)\over2L}}$. Canceling $\epsilon'$ we get an upper bound in~\eqref{eq:h1}.
\end{proof}

\section{Remarks and open problems}
\begin{itemize}
\item 
The $L/2$ in Proposition~\ref{prop:rounding} can be improved to
$O(\sqrt{L})$ using a combination of the Beck--Fiala floating colors
argument with Spencer's six deviations theorem. However, even with this
improvement, we do not see a way to prove Theorem~\ref{thm:even} with
a good value of the implicit constant.
\item 
For odd $L\geq 5$, the best upper bound we have is a tower of
exponentials of height~$L$. To that end, one colors $L$-tuples of codewords
according to the measure $\omega$ for which $\rad(x)=\mrad_{\omega}(x)$,
uses Ramsey's theorem to get a monochromatic set, 
and then proceeds similarly
to the proof of Theorem~\ref{thm:even}.
\item In the \Ldecodable/ code in Theorem~\ref{thm:simpleconstruction}, the code length is exponential in $\veps^{-1}$.
One can restrict that code to a random subset 
of $O(\veps^{-2}\log \veps^{-1})$ coordinates, 
and obtain a code of asymptotically the same size $c_L\veps^{-1}+O(1)$. 

For $L=2$ and $L=4$, the Levenshtein's code has length $O(\veps^{-1})$ and size $c_L\veps^{-1}+O(1)$.
We do not know if one can make the code in Theorem~\ref{thm:simpleconstruction}
of linear size for general~$L$.

\item It should be possible to improve Theorem~\ref{th:hilb}. We conjecture that for spherical codes, for all $L$ we have
$\maxcode_\veps(L) = O(1/\veps)$ with simplex being the optimal code.
\end{itemize}

\textbf{Acknowledgment.} We thank Alan Frieze for providing 
the reference \cite{mathworld}.

\bibliographystyle{plain}
\bibliography{listdecoding}

\end{document}